\documentclass[prd,superscriptaddress,amsfonts,amssymb,amsmath,showpacs,twocolumn]{revtex4-2}
\usepackage{bm}
\usepackage{amsfonts}
\usepackage{latexsym}
\usepackage{graphicx}
\usepackage{amsmath}
\usepackage{palatino}
\usepackage{mathpazo}
\usepackage{textcomp}
\usepackage{float}
\usepackage{booktabs}
\usepackage{dcolumn}
\usepackage{booktabs}
\usepackage{multirow}
\usepackage{hyperref}
\hypersetup{colorlinks,citecolor=blue}
\usepackage{amsmath}
\usepackage{xcolor}
\usepackage{orcidlink}
\usepackage{subcaption}
\usepackage{commath}
\def\jnl@style{\it}
\def\aaref@jnl#1{{\jnl@style#1}}
\def\aaref@jnl#1{{\jnl@style#1}}

\allowdisplaybreaks[1]

\begin{document}

\color{black}       

\title{Constraints on anisotropic properties of the universe in $f(Q, T)$ gravity theory}

\author{A. Zhadyranova\orcidlink{0000-0003-1153-3438}} 
\email[Email: ]{a.a.zhadyranova@gmail.com}
\affiliation{General and Theoretical Physics Department, L. N. Gumilyov Eurasian National University, Astana 010008, Kazakhstan.}

\author{M. Koussour\orcidlink{0000-0002-4188-0572}}
\email[Email: ]{pr.mouhssine@gmail.com}
\affiliation{Department of Physics, University of Hassan II Casablanca, Morocco.}

\author{V. Zhumabekova\orcidlink{0000-0002-7223-5373}}
\email[Email: ]{zh.venera@mail.ru}
\affiliation{Theoretical and Nuclear Physics Department, Al-Farabi Kazakh National University, Almaty, 050040, Kazakhstan.}

\author{O. Donmez\orcidlink{0000-0001-9017-2452}}
\email[Email: ]{orhan.donmez@aum.edu.kw}
\affiliation{College of Engineering and Technology, American University of the Middle East, Egaila 54200, Kuwait.}

\author{S. Muminov\orcidlink{0000-0003-2471-4836}}
\email[Email: ]{sokhibjan.muminov@gmail.com}
\affiliation{Mamun University, Bolkhovuz Street 2, Khiva 220900, Uzbekistan.}

\author{J. Rayimbaev\orcidlink{0000-0001-9293-1838}}
\email[Email: ]{javlon@astrin.uz}
\affiliation{Institute of Fundamental and Applied Research, National Research University TIIAME, Kori Niyoziy 39, Tashkent 100000, Uzbekistan.}
\affiliation{University of Tashkent for Applied Sciences, Str. Gavhar 1, Tashkent 100149, Uzbekistan.}
\affiliation{Urgench State University, Kh. Alimjan Str. 14, Urgench 221100, Uzbekistan}
\affiliation{Shahrisabz State Pedagogical Institute, Shahrisabz Str. 10, Shahrisabz 181301, Uzbekistan.}


\begin{abstract}
Motivated by anomalies in cosmic microwave background observations, we investigate the implications of $f(Q, T)$ gravity in Bianchi type-I spacetime, aiming to characterize the universe's spatially homogeneous and anisotropic properties. By using a linear combination of non-metricity $Q$ and the energy-momentum tensor trace $T$, we parametrize the deceleration parameter and derive the Hubble solution, which we then impose in the Friedmann equations of $f(Q, T)$ gravity. Bayesian analysis is employed to find the best-fit values of model parameters, with $1-\sigma$ and $2-\sigma$ contour plots illustrating the constraints from observational data, including $H(z)$ data and the Pantheon+ sample. Our analysis reveals a transition from a decelerated to an accelerated expansion phase, with the present deceleration parameter indicating an accelerating universe. The energy density gradually decreases over time, approaching zero for the present and future, indicating continuous expansion. The anisotropic pressure, initially notably negative, transitions to slightly negative values, suggesting the presence of dark energy. The evolving equation of state parameter $\omega$ exhibits behavior akin to phantom energy, influenced by spacetime anisotropy. Violations of the null energy condition and the strong energy condition imply phantom-like behavior and accelerated expansion.
\end{abstract}

\maketitle

\section{Introduction}\label{sec1}

Cosmology underwent a dramatic transformation when observational evidence from Type Ia supernovae (SNe Ia) searches \cite{Riess, Perlmutter} confirmed the accelerating expansion of the universe. In the past two decades, numerous observational results, including those from the WMAP experiment \cite{C.L., D.N.}, cosmic microwave background (CMB) radiation \cite{R.R., Z.Y.}, large-scale structure (LSS) studies \cite{T.Koivisto, S.F.}, and baryon acoustic oscillation (BAO) \cite{D.J., W.J.}, have consistently supported the observed cosmic acceleration. The leading explanation for this accelerating scenario is the presence of a dark energy (DE) component. This mysterious form of energy is commonly described by an equation of state (EoS) parameter $\omega_0 = -1.03 \pm 0.03$ in a flat universe \cite{Planck_2020}. While this value suggests a dynamical form of DE, the leading explanation remains the cosmological constant ($\Lambda$), which corresponds to $\omega_0 = -1$ in the standard $\Lambda$CDM model. DE is hypothesized to constitute approximately 68\% of the total energy density of the universe, exerting a negative pressure that drives the accelerated expansion. This concept aligns well with observational data from various cosmological probes, such as SNe Ia, CMB, and LSS surveys. The exact nature of DE remains one of the most profound mysteries in modern cosmology, with numerous theoretical models proposed to elucidate its properties and behavior. Among these models, the standard $\Lambda$CDM model is the most widely accepted by theoretical physicists for explaining DE and the late-time acceleration of the universe. However, it faces two fundamental challenges: fine-tuning and the cosmic coincidence problem \cite{S.W.,Peebles,Padmanabhan}. An alternative method for explaining the accelerating expansion of the universe, without relying on undetected DE, involves considering a more general description of the gravitational field. This approach modifies the Einstein–Hilbert action of general relativity (GR) by introducing a generic function $f(R)$, where $R$ is the Ricci scalar curvature, as initially proposed in \cite{Buchdahl/1970,Kerner/1982,Kleinert/2002}. The $f(R)$ gravity model can account for the expansion mechanism without requiring any exotic DE component \cite{Carroll/2004,Capozziello/2006}. Studies have examined the observational implications of $f(R)$ gravity models, including constraints from the solar system and the equivalence principle \cite{Tsujikawa/2008,Capozziello/2008,Starobinsky/2007}. Furthermore, viable cosmological models of $f(R)$ gravity that pass solar system tests have been proposed \cite{Nojiri/2003,Faraoni/2006,Amendola/2008}.

In the literature, several modified theories of gravity have been proposed, including $f(\mathcal{T})$ theory \cite{Paliathanasis/2016,Salako/2013,Myrzakulov/2011,Koussour_fT1}, $f(R, T)$ theory \cite{Harko/2011,Koussour_fRT,Koussour_fRT1}, $f(G)$ theory \cite{Nojiri/2008}, $f(R, G)$ theory \cite{Elizalde/2010,Bamba/2010}, and $f(R, L_m)$ theory \cite{fRL1,fRL2}. Recently, $f(Q)$ theories of gravity have garnered significant attention. Symmetric teleparallel gravity, or $ f(Q)$ gravity, was introduced by Jiménez et al. \cite{Jim/2018}, where gravitational interactions are described by the non-metricity $Q$, which geometrically represents the variation in the length of a vector during parallel transport. This is distinct from teleparallel $f(\mathcal{T})$ gravity, where gravity is described using torsion generated by tetrad fields $e^\mu_i$, replacing the metric tensor $g_{\mu \nu}$ as the primary geometric variable. In this approach, the torsion replaces curvature as the descriptor of gravitational effects. In symmetric teleparallel gravity, the covariant divergence of the metric tensor is non-zero, similar to Weyl’s theory. Numerous studies have investigated $f(Q)$ gravity \cite{Q1,Q2,Q3,Q4,Q5,Q6,Q7,Q8}, including the first cosmological solutions \cite{jimenez/2020, khyllep/2021}, geodesic deviation equations derived from its covariant formulation \cite{Avv}, and quantum cosmology for power-law models \cite{ND}. Cosmological solutions and the growth index of matter perturbations for polynomial functional forms of $f(Q)$ have also been examined \cite{WK}. Furthermore, Refs. \cite{HRK,fQL1} introduced the $f(Q,L_m)$ theory, which extends modified gravity by incorporating a non-minimal coupling between the non-metricity scalar $Q$ and the matter Lagrangian $L_m$ within the framework of the Einstein–Hilbert action.

In a recent work, Xu et al. \cite{Y.Xu} introduced an extension of $f(Q)$ gravity that involves a non-minimal coupling between the non-metricity $Q$ and the trace $T$ of the matter-energy-momentum tensor. They formulated the Lagrangian density of the gravitational field as a general function of both $Q$ and $T$, denoted as $L = f(Q, T)$. This theory resembles the $f(R, T)$ theory \cite{Harko/2011}, but instead of using the geometric sector of the Einstein–Hilbert action, it utilizes the symmetric teleparallel formulation. Like the standard couplings between the curvature and the trace of the energy-momentum tensor, the coupling between $Q$ and $T$ in the $f(Q, T)$ theory also causes the energy-momentum tensor to be non-conserved. This non-conservation has important physical consequences, including substantial changes in the thermodynamics of the universe, similar to those seen in the $f(R, T)$ theory \cite{Harko/2011}. Further, the non-geodesic motion of test particles leads to the emergence of an extra force. Koussour et al. \cite{K6,K7} specifically examined the late-time accelerated expansion of the universe using observational constraints. Additionally, significant research has been conducted on topics such as baryogenesis \cite{Bhattacharjee}, cosmological inflation \cite{Shiravand}, and cosmological perturbations \cite{Najera}. However, the astrophysical implications of $f(Q,T)$ gravity have not been extensively explored. Tayde et al. \cite{Tayde1} investigated static spherically symmetric wormhole solutions in $f(Q,T)$ gravity, considering both linear and non-linear models under various equations of state. Furthermore, Pradhan et al. \cite{Sneha2} examined the thin-shell gravastar model within the framework of $f(Q, T)$ gravity.. Recently, Bourakadi et al. \cite{Bourakadi} explored constant-roll inflation and the formation of primordial black holes within the framework of $f(Q, T)$ gravity.

Contrary to popular belief, which holds that the universe is homogeneous and isotropic, there is evidence that anisotropies may have existed in the early past and may reappear in the future. Small-scale anisotropies in the CMB were discovered by the cosmic background explorer in 1996 \cite{Bennett_1996}. This idea of anisotropic spacetime geometry is substantially supported by observational data from experiments like the cosmic background imager \cite{Mason_2003} and the WMAP \cite{Hinshaw_2013}. Moreover, recent developments indicate that the cosmos is expanding anisotropically, as indicated by differences in the intensities of microwaves received from various directions \cite{Migkas_2020}. An efficient framework for characterizing the homogeneous and anisotropic properties of spacetime is presented by Bianchi-type cosmology. In the literature, numerous models inspired by Bianchi cosmology have been explored \cite{Amirhashchi_2017,Amirhashchi_2018,Amirhashchi_2020,Shamir_2015,Rodrigues_2012}. Recent studies have focused on the isotropization mechanism in anisotropic Bianchi type-I cosmology, particularly through a polynomial $f(Q)$ model \cite{iso}. Loo et al. \cite{Loo_2023} investigated the dynamics of an anisotropic universe in the context of $f(Q, T)$ gravity, while Narawade et al. \cite{Narawade_2023} examined observational constraints on the hybrid scale factor in anisotropic spacetimes under $f(Q,T)$ gravity. In this work, we use the locally rotationally symmetric (LRS) Bianchi type-I metric, which is essentially an extension of the isotropic case and has a strong resemblance to the Friedmann–Lema\^itre–Robertson–Walker (FLRW) metric. This work differs from previous studies by examining the anisotropic dynamics of the universe in $f(Q,T)$ gravity within a Bianchi type-I spacetime. Specifically, we consider a linear combination of $Q$ and $T$ in the form $f(Q,T) = Q + b T$ \cite{Bekkhozhayev}, motivated by its simplicity and ability to capture deviations from GR while still allowing for analytic solutions in an anisotropic spacetime.

The present study is organized as follows: Sec. \ref{sec2} outlines the basic formalism of $f(Q, T)$ gravity. In Sec. \ref{sec3}, our focus is on the anisotropic Bianchi type-I cosmological model and the equations of motion in the context of $f(Q, T)$ gravity. Sec. \ref{sec4} is dedicated to exploring a particular functional form of $f(Q, T)$ and its cosmological solutions for $f(Q, T)$ gravity, using parameterization of the deceleration parameter. In Sec. \ref{sec5}, we employ observational data from $H(z)$ data and Pantheon+ sample to determine the model parameters. In Sec. \ref{sec6}, we discuss the cosmological implications of the model and validate its consistency with the energy conditions. Lastly, in Sec. \ref{sec7}, we review and summarize our results.

\section{Basic formalism of $f(Q,T)$ gravity} \label{sec2} 

The action principle for the $f(Q,T)$ gravity model, as proposed by Xu et al. \cite{Y.Xu}, is defined using the non-metricity scalar $Q$ and the trace of the energy-momentum tensor $T$ as
\begin{equation}  \label{1}
S=\int \sqrt{-g}\left[ \frac{1}{16\pi }f(Q,T)+L_{m}\right] d^{4}x,
\end{equation}
where $f(Q,T)$ is an arbitrary function of $Q$ and $T$, and $g$ is the determinant of the metric tensor $g_{\mu \nu }$. 

The non-metricity scalar $Q$ is defined in terms of the metric tensor $g_{\mu \nu}$ and the disformation tensor $L_{\alpha \gamma }^{\beta }$ as follows:
\begin{equation}
Q\equiv -g^{\mu \nu }(L_{\,\,\,\alpha \mu }^{\beta }L_{\,\,\,\nu \beta
}^{\alpha }-L_{\,\,\,\alpha \beta }^{\beta }L_{\,\,\,\mu \nu }^{\alpha }),
\label{2}
\end{equation}%
where 
\begin{equation}
L_{\alpha \gamma }^{\beta }=\frac{1}{2}g^{\beta \eta }\left( Q_{\gamma
\alpha \eta }+Q_{\alpha \eta \gamma }-Q_{\eta \alpha \gamma }\right) ={%
L^{\beta }}_{\gamma \alpha }.  \label{3}
\end{equation}
and
\begin{equation}
Q_{\gamma \mu \nu }=-\nabla _{\gamma }g_{\mu \nu }=-\partial _{\gamma
}g_{\mu \nu }+g_{\nu \sigma }\widetilde{\Gamma }{^{\sigma }}_{\mu \gamma
}+g_{\sigma \mu }\widetilde{\Gamma }{^{\sigma }}_{\nu \gamma }.  \label{4}
\end{equation}%

Here, $\widetilde{\Gamma }{^{\gamma }}_{\mu \nu }=L^{\gamma }{}_{\mu \nu}+\Gamma {^{\gamma }}_{\mu \nu }$ represents the Weyl connection, where $\Gamma {^{\gamma }}_{\mu \nu }$ denotes the well-known Levi-Civita connection associated with the metric. In addition, the trace of the non-metricity tensor is given by
\begin{equation}
Q_{\beta }=g^{\mu \nu }Q_{\beta \mu \nu },\qquad \widetilde{Q}_{\beta
}=g^{\mu \nu }Q_{\mu \beta \nu }.  \label{5}
\end{equation}%

Further, we introduce the superpotential tensor, also referred to as the non-metricity conjugate. It is defined as follows:
\begin{eqnarray}
\hspace{-0.5cm} &&P_{\ \ \mu \nu }^{\beta }\equiv \frac{1}{4}\bigg[-Q_{\ \
\mu \nu }^{\beta }+2Q_{\left( \mu \ \ \ \nu \right) }^{\ \ \ \beta
}+Q^{\beta }g_{\mu \nu }-\widetilde{Q}^{\beta }g_{\mu \nu }  \notag \\
\hspace{-0.5cm} &&-\delta _{\ \ (\mu }^{\beta }Q_{\nu )}\bigg]=-\frac{1}{2}%
L_{\ \ \mu \nu }^{\beta }+\frac{1}{4}\left( Q^{\beta }-\widetilde{Q}^{\beta
}\right) g_{\mu \nu }-\frac{1}{4}\delta _{\ \ (\mu }^{\beta }Q_{\nu )}.\quad
\quad   \label{6}
\end{eqnarray}

Hence, the non-metricity scalar is defined as \cite{Jim/2018,jimenez/2020}, 
\begin{eqnarray}
&&Q=-Q_{\beta \mu \nu }P^{\beta \mu \nu }=-\frac{1}{4}\big(-Q^{\beta \nu
\rho }Q_{\beta \nu \rho }+2Q^{\beta \nu \rho }Q_{\rho \beta \nu }  \notag \\
&&-2Q^{\rho }\tilde{Q}_{\rho }+Q^{\rho }Q_{\rho }\big).  \label{7}
\end{eqnarray}

The field equation of $f(Q,T)$ gravity is derived by varying the action \eqref{1} with respect to the metric component $g_{\mu \nu}$,
\begin{multline}
\frac{2}{\sqrt{-g}}\nabla _{\beta }(f_{Q}\sqrt{-g}P_{\,\,\,\,\mu \nu
}^{\beta })-\frac{1}{2}fg_{\mu \nu }+f_{T}(T_{\mu \nu }+\Theta _{\mu \nu })
\label{11} \\
+f_{Q}(P_{\mu \beta \alpha }Q_{\nu }^{\,\,\,\beta \alpha }-2Q_{\,\,\,\mu
}^{\beta \alpha }P_{\beta \alpha \nu })=8\pi T_{\mu \nu }.
\end{multline}
where $f\equiv f\left(Q,T\right)$, $f_Q \equiv \frac{\partial f}{\partial Q}$, $f_T \equiv \frac{\partial f}{\partial T}$, and $T_{\mu \nu }$ is the energy-momentum tensor derived from the matter Lagrangian and is defined as
\begin{equation}
T_{\mu \nu }=-\frac{2}{\sqrt{-g}}\dfrac{\delta (\sqrt{-g}L_{m})}{\delta
g^{\mu \nu }}.
\end{equation}

Furthermore, we have
\begin{equation}  \label{10}
\frac{\delta \,g^{\,\mu \nu }\,T_{\,\mu \nu }}{\delta \,g^{\,\alpha \,\beta }%
}=T_{\,\alpha \beta }+\Theta _{\,\alpha \,\beta }\,,
\end{equation}
where $\Theta _{\mu \nu }=g^{\alpha \beta }\frac{\delta T_{\alpha \beta }}{\delta
g^{\mu \nu }}$. In addition, it is crucial to note that in $f(Q,T)$ gravity, the divergence of the energy-momentum tensor can be expressed as $\mathcal{D}_{\mu }T_{\ \ \nu }^{\mu }=B_{\nu }\neq 0$, where $\mathcal{D}_{\mu }$ denotes the covariant derivative with respect to the connection associated with the non-metricity $Q$. The term $B_{\nu}$ is the non-conservation vector, which depends on $Q$, $T$, and the thermodynamic properties of the system. Therefore, in this framework, the energy-momentum tensor is not conserved.

\section{Bianchi type-I metric and motion equations in $f(Q,T)$ gravity}\label{sec3} 

In the context of cosmology, the Bianchi type-I metric plays a significant role in describing an anisotropic universe. Unlike the isotropic models, where isotropy implies a uniform expansion in all directions, the Bianchi type-I metric allows for different scale factors along each spatial direction. This anisotropy introduces complexities into the dynamics of the universe, leading to distinct observational signatures compared to isotropic models. Understanding the behavior of cosmological models under this metric is crucial for accurately modeling the evolution of our universe and interpreting observational data. The metric is diagonal and can be written as
\begin{equation} \label{B1}
ds^{2}=-dt^{2}+A^{2}(t)dx^{2}+B^{2}(t)(dy^{2}+dz^{2}),
\end{equation}
where $A\left( t\right)$ and $B\left( t\right)$ are the scale factors corresponding to each spatial dimension, and $t$ is the cosmic time. Here, we assume the symmetry between $y$ and $z$ to simplify the model, as this assumption is often sufficient to capture the effects of anisotropy while preserving analytical traceability \cite{Q4}. In particular, the standard flat FLRW cosmology is derived when the scale factors $A(t)$ and $B(t)$ are equal to the scale factor $a(t)$.

The non-metricity scalar associated with the Bianchi type-I metric can then be expressed as \cite{Loo_2023}:
\begin{equation}
Q=-6(2H-H_{y})H_{y}\,,  \label{Q}
\end{equation}
where $H_{x}=\frac{\dot{A}}{A}$, $H_{y}=\frac{\dot{B}}{B}$, and $H_{z}=H_{y}$ denote directional Hubble parameters. Moreover, the average Hubble parameter, which characterizes the rate of expansion of the universe, can be computed as
\begin{equation}
H=\frac{\dot{a}}{a}=\frac{1}{3}\left( H_{x}+2H_{y}\right) \,.
\label{H}
\end{equation}

By examining Eq. (\ref{Q}), we observe that when $H_{x}=H_{y}=H$, it simplifies to $Q=-6H^{2}$, which corresponds to the isotropic FLRW case. 

The anisotropy parameter quantifies the degree of deviation from isotropy in a cosmological model. It measures how much the expansion rate of the universe differs along different spatial directions. It is defined as
\begin{equation}
\Delta =\frac{1}{3}\sum_{i=1}^{3}\left( \frac{H_{i}-H}{H}\right) ^{2}=\frac{2%
}{9H^{2}}\left( H_{x}-H_{y}\right) ^{2}.  \label{eqn25}
\end{equation}

In addition, the expansion scalar $\theta$ and the shear scalar $\sigma$ of the fluid are defined as follows:
\begin{equation}
\theta=H_{x}+2H_{y}=3H,\quad \sigma=\frac{1}{%
\sqrt{3}}\left( H_{x}-H_{y}\right) .  \label{eqn26}
\end{equation}

In cosmology, a perfect fluid is a theoretical model often used to describe a continuous distribution of matter exhibiting certain idealized properties. Its simplicity and ability to describe large-scale properties of matter make it a valuable tool in studying the dynamics of the universe. The energy-momentum tensor associated with the metric (\ref{B1}) is described by the standard form for a perfect fluid:
\begin{equation}
T_{\mu\nu}=(\rho+p)u_\mu u_\nu + p g_{\mu\nu}.
\end{equation}

Here $\rho$ is the energy density, $p$ is the anisotropic pressure, and $u^\mu=(1,0,0,0)$ is the four-velocity of the fluid element. In addition, we assume the matter Lagrangian to be $L_{m}=p$, which results in $\Theta _{\mu \nu }=pg_{\mu \nu }-2T_{\mu \nu }$.

The generalized Friedmann equations, which describe the dynamics of the universe in $f(Q, T)$ gravity, are expressed as \cite{Loo_2023,Narawade_2023},
\begin{align}
(8\pi +f_{T})\rho  &+f_{T}p= \frac{f}{2}+6f_{Q}(2H-H_{y})H_{y}\,,  \label{rho} \\
8\pi p&= -\frac{f}{2}-\frac{\partial }{\partial t}\left[ 2f_{Q}H_{y}\right] -6f_{Q}H_{y}H,  \label{p_x} \\
8\pi p&= -\frac{f}{2}-\frac{\partial }{\partial t}\left[ f_{Q}(3H-H_{y})\right] -3f_{Q}(3H-H_{y})H\,.  \label{p_y}
\end{align}

Through algebraic manipulations of Eqs. (\ref{rho}) to (\ref{p_y}), we can simplify the field equations to
\begin{eqnarray}
\label{rho-2} 
8\pi \rho &= & \frac{f}{2} + \frac{6f_{Q}}{8\pi + f_{T}} \left[ 8\pi (2H - H_{y})H_{y} + f_{T}H^{2} \right] \\ 
& +& \frac{2f_{T}}{8\pi + f_{T}} \frac{\partial}{\partial t} \left[ f_{Q}H \right], \nonumber \\
8\pi p&=& -\frac{f}{2}-2\frac{\partial }{\partial t}\left[ f_{Q}H\right]
-6f_{Q}H^{2}\,.  
\label{p-2}
\end{eqnarray}

\section{Cosmological solutions} \label{sec4}

Recently, Xu et al. \cite{Y.Xu} considered a linear form for the function $f(Q, T)$, in which the cosmological evolution follows a de Sitter type, resulting in the universe expanding exponentially. Here, to investigate the anisotropic dynamics of the universe, we consider the simple linear functional form of $f(Q, T)$ gravity given by \cite{Narawade_2023}
\begin{equation} 
f\left( Q,T\right) =Q+b T,    
\end{equation}
where $b$ is a free parameter, and the scenario equivalent to GR is retained for $b=0$. In this scenario, we have $f_{Q}=1$ and $f_{T}=b$. Therefore, Eqs. (\ref{rho-2}) and (\ref{p-2}) can be rewritten as
\begin{eqnarray}
\label{rho-3}
\rho &=& \frac{3H^{2}+\overset{.}{H}}{2b+8\pi }-\frac{3(H-H_{y})^{2}+\overset%
{.}{H}}{b+8\pi }~, \\
p &=& -\frac{3H^{2}+\overset{.}{H}}{2b+8\pi }-\frac{3(H-H_{y})^{2}+\overset{.%
}{H}}{b+8\pi },
\label{p-3}
\end{eqnarray}
where the dot represents the derivative with respect to cosmic time $t$. Further, we assume a physical condition where the shear scalar is proportional to the expansion scalar ($\sigma ^{2}\propto \theta ^{2}$), leading to the relation $A = B^n$, where $n$ is a real number. This physical law is based on observations of the velocity-redshift relation for extragalactic sources, suggesting that the Hubble expansion of the universe may achieve isotropy when $\frac{\sigma }{\theta }=constant$ \cite{Collins}. This condition has been used in several instances in the literature \cite{Loo_2023,Narawade_2023,Collins,Rodrigues,Bishi}. In terms of the directional Hubble parameter, this condition can be expressed as $H_{x}=n H_{y}, n\neq (0,1)$. For $n=1$, the isotropic flat FLRW cosmology is recovered. Thus,  the average Hubble parameter is expressed as
\begin{equation} \label{Hy}
H=\frac{\left( n +2\right) }{3}H_{y}.
\end{equation}

Now, we have a system of two equations as described in Eqs. (\ref{rho-3}) and (\ref{p-3}), which involve three unknowns: $H$, $p$, and $\rho$. Therefore, to obtain a complete solution for the system, an additional plausible condition is necessary to determine the unique solution for this system of equations.

In this study, the focus is on a specific form of parametrization of the deceleration parameter $q$ as defined by Koussour et al. \cite{DP}, which is represented by
\begin{equation} \label{DP}
q=-\frac{\overset{..}{a}}{aH^{2}}=-1+\frac{\alpha }{1+\beta a^{3}}
\end{equation}
Where $\alpha$ and $\beta$ are positive constants, and $a$ is the scale factor of the universe, defined in terms of redshift $z$ as $a=(1+z)^{-1}$. The sign of the deceleration parameter indicates the direction of the universe's expansion. For $q > 0$, the universe experiences deceleration, meaning its expansion rate decreases over time. In contrast, for $q < 0$, the universe undergoes acceleration, indicating an increasing expansion rate. It is important to note that current observations, such as SNe Ia and CMB \cite{Riess, Perlmutter,R.R., Z.Y.}, tend to favor accelerating models characterized by $q < 0$. From the above expression, we can see that at present ($z=0$), the current value of the deceleration parameter $q_{0}$ is given by $q_{0}=-1+\frac{\alpha }{1+\beta }$. This implies that the present-day acceleration or deceleration of the Universe depends on the values of $\alpha $ and $\beta $. If $\alpha <1+\beta $, the Universe is currently accelerating; if $\alpha >1+\beta $, the Universe is currently decelerating; and if $\alpha =1+\beta $, the universe is currently coasting.

The following equation establishes the relationship between the deceleration parameter and the Hubble parameter:
\begin{equation}
H\left( z\right) =H_{0}\exp \left( \int_{0}^{z}\frac{1+q\left( z\right) }{%
\left( 1+z\right) }dz\right) .  \label{Hq}
\end{equation}

By substituting Eq. (\ref{DP}) into Eq. (\ref{Hq}), we derive the expression for $H(z)$ as follows:
\begin{equation}
H\left( z\right) =H_{0} \left(\frac{(1+z)^3+\beta}{1+\beta}\right)^{\alpha /3},
\label{Hz}
\end{equation}%
where $H_0$ denotes the present value of the Hubble parameter (at $z = 0$). The time derivative of the Hubble parameter can be expressed as follows: $\overset{.}{H}=-\left( 1+z\right) H\left( z\right) \frac{dH\left( z\right) }{dz}$. Using Eq. (\ref{Hz}), we have
\begin{equation}
\overset{.}{H}=-\frac{\alpha  H_{0}^2 (1+z)^3}{1+\beta} \left(\frac{(1+z)^3+\beta}{1+\beta}\right)^{\frac{2 \alpha }{3}-1}.
\end{equation}

By substituting Eq. (\ref{Hz}) into Eqs. (\ref{eqn25}) and (\ref{eqn26}), we derive the expressions for the anisotropy parameter, scalar expansion, and shear scalar as follows:
\begin{eqnarray}
\Delta &=& \frac{2 (n-1)^2}{(n+2)^2} , \\
\theta &=& 3 H_{0} \left(\frac{(1+z)^3+\beta}{1+\beta}\right)^{\alpha /3}, \\
\sigma &=& \frac{\sqrt{3} H_{0} (n-1)}{n+2} \left(\frac{(1+z)^3+\beta}{1+\beta}\right)^{\alpha /3}.
\end{eqnarray}
respectively. From the above expressions, it is evident that the anisotropy parameter remains constant, indicating uniform anisotropy throughout the evolution of the universe. The scalar expansion and shear scalar diverge during the early stages ($z>>1$) of the universe and approach zero as time progresses toward infinity ($z \rightarrow -1$). This indicates that the universe initially undergoes a phase of infinite expansion rate, eventually transitioning to a constant expansion rate in later epochs. 

\section{Observational data}\label{sec5}

In this section, a statistical analysis is performed to compare the predictions of the theoretical model with observational data. The goal is to establish constraints on the free parameters of the model, namely $H_0$, $\alpha$, and $\beta$. The analysis employs a sample of Cosmic Chronometers (CC), consisting of 31 measurements, along with the Pantheon+ sample, which includes 1701 data points. We use the emcee Python package \cite{Mackey}, which implements the affine-invariant ensemble sampler for Markov Chain Monte Carlo (MCMC) simulations, a widely used tool in Bayesian methods in cosmology \cite{BS}. The MCMC sampler is used to estimate the posterior distribution of the model parameters. The likelihood function is constructed using observational data, and the posterior distribution is computed through multiple iterations of MCMC sampling. Our MCMC analysis uses 100 walkers and 1000 steps to obtain the fitting results. The likelihood function is given by \cite{dat1,dat2}
\begin{equation}
\mathcal{L}\propto \exp(-\chi^2 / 2).
\end{equation}
where $\chi^2$ represents the pseudo chi-squared function \cite{BS}. Further details about the construction of the $\chi^2$ function for different data samples are provided in the following subsections.

\subsection{Observational $H(z)$ data}

Cosmic chronometers are a method for estimating the Hubble parameter by comparing the relative ages of passively evolving galaxies. They are identifiable by specific features in their spectra and color profiles \cite{R35}. Cosmic chronometers use the estimated ages of galaxies at various redshifts to obtain their data. For our analysis, we incorporate a collection of 31 independent measurements of \( H(z) \) spanning the redshift range \( 0.07 \leq z \leq 2.41 \) \cite{R36}. These \( H(z) \) measurements are obtained using the relationship \( H(z) = -\frac{1}{1+z} \frac{dz}{dt} \). In this context, \( \frac{dz}{dt} \) is estimated by \( \frac{\Delta z}{\Delta t} \), where \( \Delta z \) and \( \Delta t \) denote the change in redshift and the corresponding change in age between two galaxies. The corresponding \(\chi^2\) function is given by:
\begin{equation}\label{4a}
\chi_{Hz}^{2}=\sum\limits_{k=1}^{31}\frac{[H_{th}(z_{k})-H_{obs,k}]^{2}}{\sigma _{H,k}^{2}}.  
\end{equation}
Here, \( H_{obs,k} \) represents the observed Hubble value, while \( H_{th}(z_{k}) \) denotes the theoretical value of \( H(z) \) at redshift \( z_{k} \). The standard error is given by \( \sigma_{H,k} \).

\subsection{Pantheon+ SNe Ia sample}

The Pantheon+ sample spans a wide range of redshifts from 0.001 to 2.3, incorporating the latest observational data and surpassing previous collections of SNe Ia. SNe Ia are known for their consistent brightness, making them reliable standard candles for measuring relative distances using the distance modulus technique. Over the past two decades, several compilations of SNe Ia data have been introduced, including Union \cite{R37}, Union2 \cite{R38}, Union2.1 \cite{R39}, JLA \cite{R40}, Pantheon \cite{R41}, and the latest addition, Pantheon+ \cite{R42}. The corresponding \(\chi^2\) function is given by:
\begin{equation}\label{4b}
\chi^2_{SNe}= D^T C^{-1}_{SN} D,
\end{equation}
In this context, \( C_{SN} \) \cite{R42} denotes the covariance matrix linked with the Pantheon+ samples, which includes both statistical and systematic uncertainties. Also, the vector \( D \) is defined as \( D = m_{Bi} - M - \mu^{th}(z_i) \), where \( m_{Bi} \) and \( M \) represent the apparent magnitude and absolute magnitude, respectively. Furthermore, \( \mu^{th}(z_i) \) represents the distance modulus of the assumed theoretical model, and it can be expressed as:
\begin{equation}\label{4c}
\mu^{th}(z_i)= 5log_{10} \left[ \frac{D_{L}(z_i)}{1 Mpc}  \right]+25, 
\end{equation}
where \(d_{L}(z)\) represents the luminosity distance in the assumed theoretical model, and it can be expressed as follows:
\begin{equation}\label{4d}
d_{L}(z)= c(1+z) \int_{0}^{z} \frac{ dx}{H(x,\theta)}
\end{equation}
where \(\theta\) represents the parameter space of the assumed model. In contrast to the Pantheon dataset, the Pantheon+ compilation effectively resolves the degeneracy between the parameters \( H_0 \) and \( M \) by redefining the vector \( D \) as follows:
\begin{equation}\label{4e}
\bar{D} = \begin{cases}
     m_{Bi}-M-\mu_i^{Ceph} & i \in \text{Cepheid hosts} \\
     m_{Bi}-M-\mu^{th}(z_i) & \text{otherwise}
    \end{cases}   
\end{equation}
Here, \(\mu_i^{Ceph}\) is independently estimated using Cepheid calibrators. Consequently, the equation \(\chi^2_{SNe}= \bar{D}^T C^{-1}_{SN} \bar{D} \) is obtained.\\

In addition, we use the total \(\chi^2_{total}\) to obtain joint constraints for the parameters $H_0$, $\alpha$, and $\beta$ from the $H(z)$ and Pantheon+ samples. Thus, the relevant chi-square functions are defined as 
\begin{equation}
\chi^2_{total}= \chi^2_{Hz}+\chi^2_{SNe}    
\end{equation}

\begin{figure}[H]
\centerline{\includegraphics[scale=0.6]{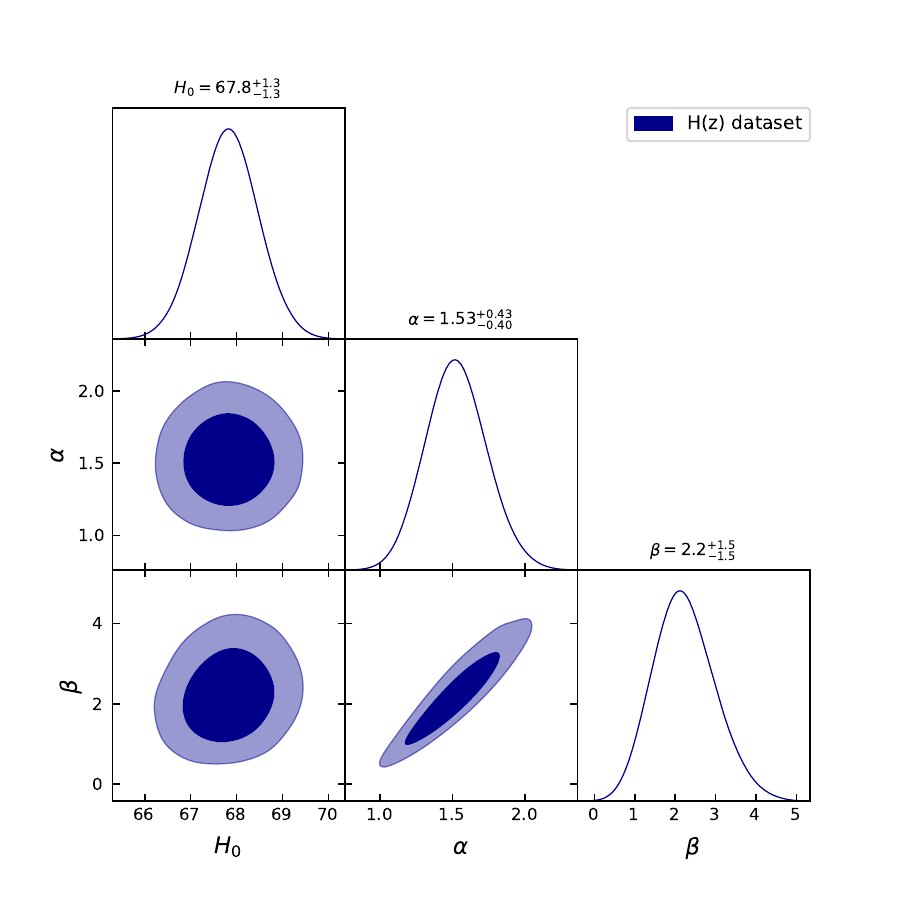}}
\caption{Confidence intervals for model parameters using the $H(z)$ dataset: 1-$\sigma$ and 2-$\sigma$ levels.}
\label{F_CC}
\end{figure}

\begin{figure}[H]
\centerline{\includegraphics[scale=0.6]{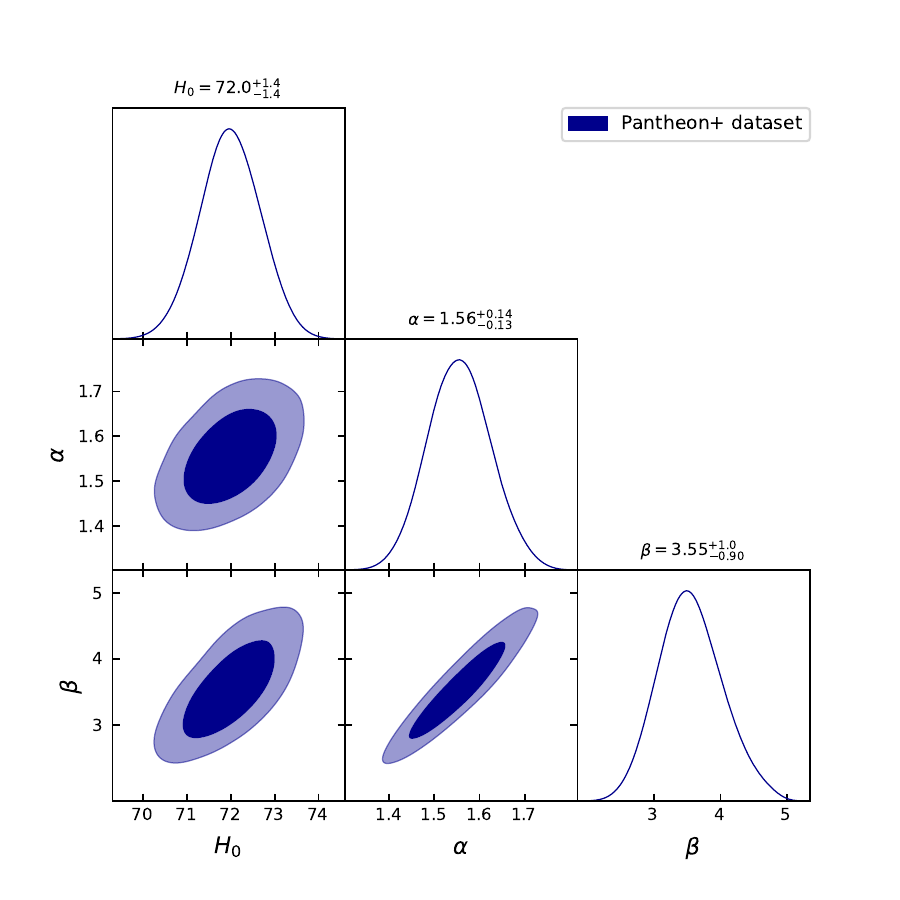}}
\caption{Confidence intervals for model parameters using the Pantheon+ SNe Ia sample: 1-$\sigma$ and 2-$\sigma$ levels.}
\label{F_SNe}
\end{figure}

\begin{figure}[H]
\centerline{\includegraphics[scale=0.6]{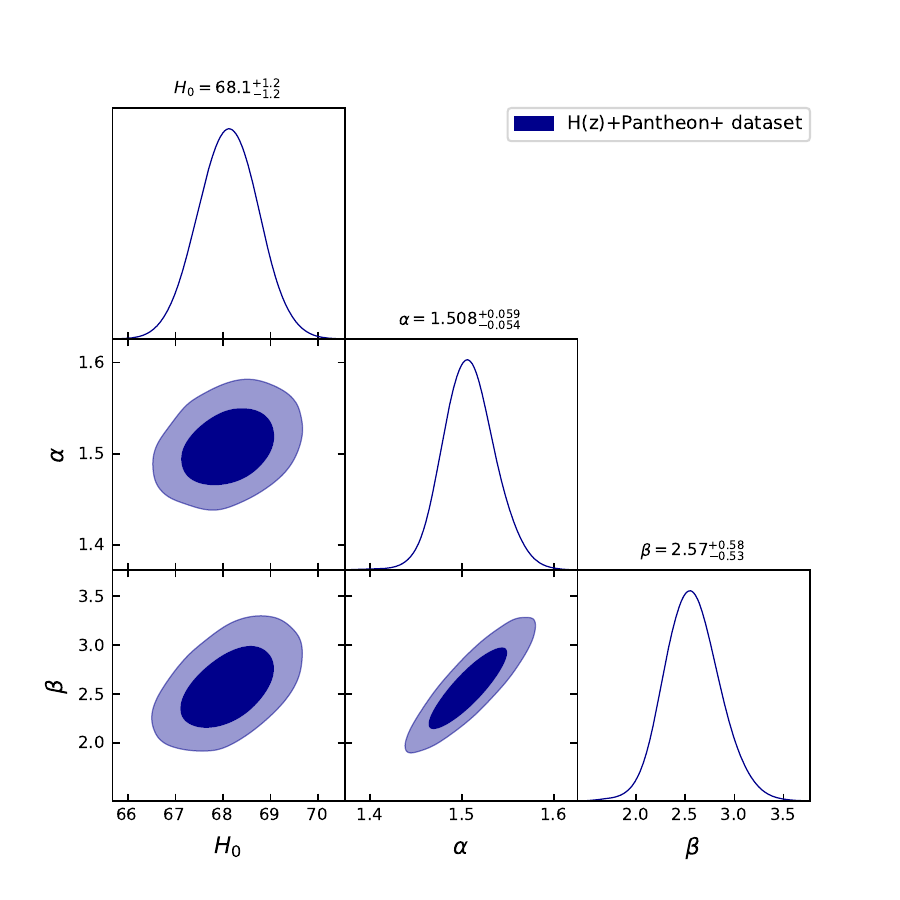}}
\caption{Confidence intervals for model parameters using the combined dataset: 1-$\sigma$ and 2-$\sigma$ levels.}
\label{F_CC+SNe}
\end{figure}

In our analysis, we use Gaussian priors as follows: \([50,100]\) for \(H_0\), \([0,10]\) for \(\alpha\), and \([0,10]\) for \(\beta\). The corresponding $1-\sigma$ and $2-\sigma$ contour plots for the $H(z)$ data, Pantheon+ sample, and the combined observational data are displayed in Figs. \ref{F_CC}, \ref{F_SNe}, and \ref{F_CC+SNe}, respectively. The results with a \(68\%\) confidence limit are as follows: \(H_0=67.8^{+1.3}_{-1.3}\), \(\alpha=1.53^{+0.43}_{-0.40}\), and \(\beta=2.2^{+1.5}_{-1.5}\) for \(H(z)\) data; \(H_0=72.0^{+1.4}_{-1.4}\), \(\alpha=1.56^{+0.14}_{-0.13}\), and \(\beta=3.55^{+1.0}_{-0.90}\) for Pantheon+ data; \(H_0=68.1^{+1.2}_{-1.2}\), \(\alpha=1.508^{+0.059}_{-0.054}\), and \(\beta=2.57^{+0.58}_{-0.53}\) for the combined data.

\section{Cosmological implications of the model}
\label{sec6}

\subsection{Cosmological parameters}

\begin{figure}[h]
\centerline{\includegraphics[scale=0.7]{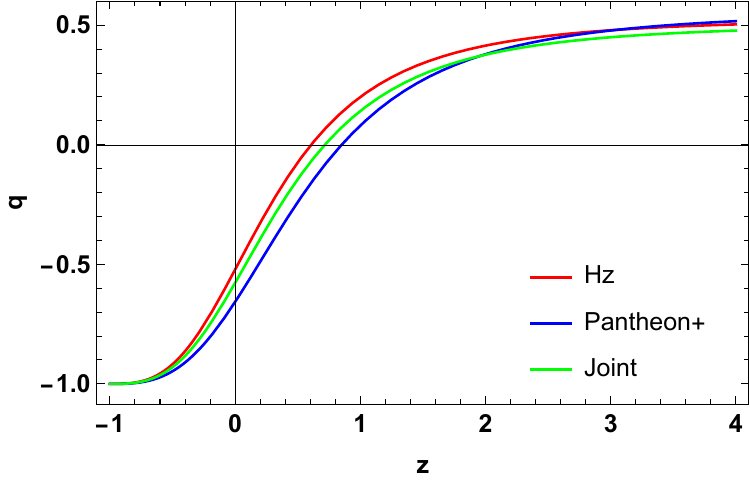}}
\caption{Variation of the deceleration parameter as a function of redshift $z$.}
\label{F_q}
\end{figure}

The deceleration parameter is a crucial metric for understanding the universe's expansion phases. Fig. \ref{F_q} shows that the model transitions from a decelerated epoch to a de-sitter-type accelerated expansion phase. Specifically, the transition redshift is $z_t = 0.61$ for the $H(z)$ data, $z_t = 0.85$ for the Pantheon+ samples, and $z_t = 0.72$ for the combined data \cite{Jesus,Garza,Mamon2,Farooq1}. The present value of the deceleration parameter is $q(z = 0) = q_0 = -0.52$, $q_0 =-0.65$, and $q_0 =-0.57$, respectively, at a 68\% confidence limit, aligning well with observed values \cite{Hernandez,Basilakos,Koussour1,Koussour2,Koussour3,Koussour4}. This consistency with observational data supports the model's validity and its effectiveness in describing the universe's expansion history. The uniform transition redshift across different datasets further highlights the robustness of the model's predictions.

\begin{figure}[h]
\centerline{\includegraphics[scale=0.7]{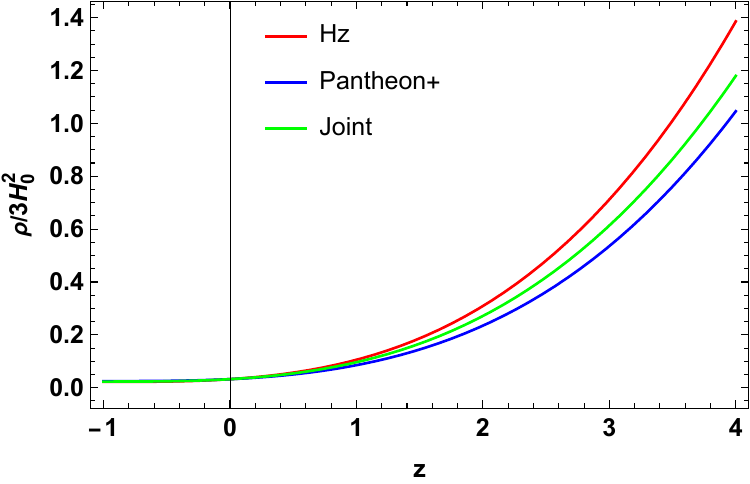}}
\caption{Variation of the energy density as a function of redshift $z$ with $b= -0.1$ and $n = 0.087$ \cite{Bekkhozhayev}.}
\label{F_rho}
\end{figure}

\begin{figure}[h]
\centerline{\includegraphics[scale=0.7]{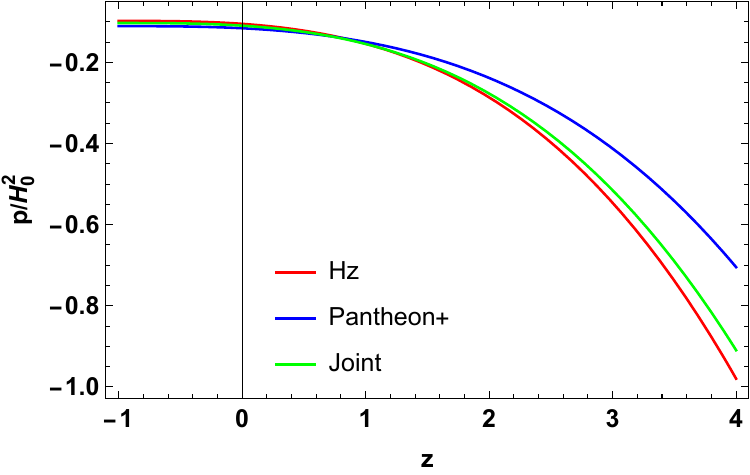}}
\caption{Variation of the pressure as a function of redshift $z$ with $b= -0.1$ and $n = 0.087$ \cite{Bekkhozhayev}.}
\label{F_p}
\end{figure}

From Fig. \ref{F_rho}, the energy density demonstrates a compelling trend across all constrained values of the model parameters. It starts with a significant initial magnitude but gradually diminishes over time, ultimately converging towards zero for the present ($z=0$) and future ($z \rightarrow -1$). This striking behavior strongly suggests the continuous expansion of the universe. As illustrated in Fig. \ref{F_p}, the pressure initially shows significantly negative values, which gradually evolve toward less negative values over time. This gradual shift in the pressure profile reflects the influence of modified gravity effects within the model, without requiring an explicit DE component. This behavior aligns with the modified gravity framework, where late-time cosmic acceleration is driven by geometric contributions, rather than the introduction of a separate DE term.

\begin{figure}[h]
\centerline{\includegraphics[scale=0.7]{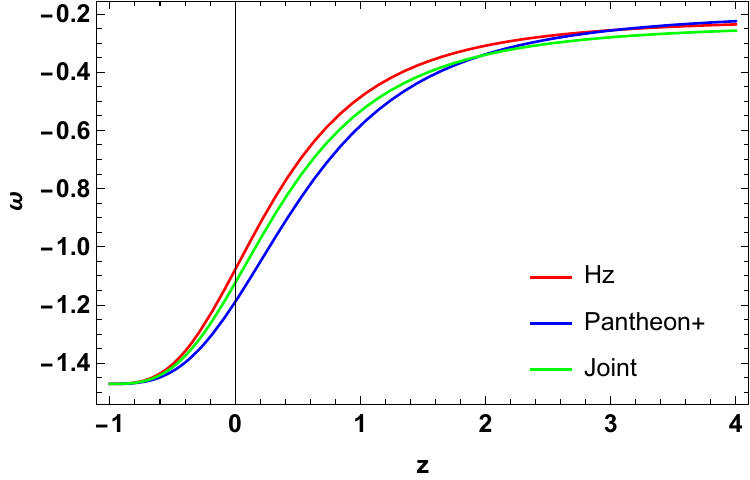}}
\caption{Variation of the EoS parameter as a function of redshift $z$ with $b= -0.1$ and $n = 0.087$ \cite{Bekkhozhayev}.}
\label{F_EoS}
\end{figure}

The equation of state (EoS) parameter in cosmology describes the relationship between pressure and energy density for a given substance or component of the universe. It is denoted by the symbol $\omega$ and is defined as the ratio of anisotropic pressure $p$ to energy density $\rho$:
\begin{equation}
\omega = \frac{p}{\rho}.   
\end{equation}

This parameter plays a crucial role in determining the behavior of different components of the universe, such as DE, dark matter, and radiation, as they evolve over time. The value of the EoS parameter can vary depending on the substance: For ordinary matter (baryonic matter and non-relativistic matter), the EoS parameter is approximately $\omega = 0$, indicating that the pressure is negligible compared to the energy density. For radiation, including photons and relativistic particles, the EoS parameter is $\omega = \frac{1}{3}$, indicating that the pressure is one-third of the energy density. For DE, which is thought to be driving the accelerated expansion of the universe, the EoS parameter is typically represented by $\omega \approx -1$. This value is consistent with a cosmological constant ($\Lambda$) or vacuum energy, leading to a constant energy density and negative pressure. In addition to the values mentioned earlier, there are two other important categories in cosmology for the EoS parameter:
\begin{itemize}
\item Quintessence \cite{RP,M.T.,LX}: is a hypothetical form of DE postulated to explain the accelerating expansion of the universe. It is characterized by a dynamic EoS parameter $\omega$ that evolves over time. Typically, quintessence models have $-1 < \omega < -\frac{1}{3}$, allowing for a range of behaviors that can mimic both cosmological constant-like behavior ($\omega \approx -1$) and more exotic forms of DE.
\item Phantom \cite{KI,DB,DB-2}: is another theoretical form of DE with an EoS parameter $\omega < -1$. This implies that the energy density of phantom energy increases as the universe expands, leading to a "big rip" scenario where the universe is ultimately torn apart. The concept of phantom DE is highly speculative and is not supported by current observational data.
\end{itemize}

The EoS parameter is often dynamic, particularly for DE, allowing for transitions between different epochs. A notable example is a fluid with $\omega > -1$ at certain times and $\omega < -1$ at others, combining quintessence-like and phantom-like behaviors. Such a transition can be seen in Fig. \ref{F_EoS}, where the EoS parameter evolves from the quintessence region ($ \omega > -1$) to the phantom regime ($ \omega < -1 $) as the universe evolves. This dynamic evolution indicates that the EoS parameter is not static, but rather evolves with time, a behavior characteristic of models that allow for EoS crossing the phantom divide line ($\omega= -1$). It is important to clarify that while quintessence and phantom DE are commonly explored within cosmological models, the results discussed here pertain specifically to our model within the framework of $f(Q,T)$ modified gravity. Therefore, the crossing of the phantom divide line in our context does not imply the existence of a traditional DE fluid, but rather emerges from the geometric contributions of the modified gravity model itself. Observational data, such as from SNe Ia and the CMB, are used to constrain the possible values of $\omega$ and understand the nature of DE. The WMAP9 data \cite{Hinshaw_2013}, combining measurements from the Hubble parameter ($H_0$), SNe Ia, CMB, and BAO, suggests $\omega_0 = -1.084 \pm 0.063$. In contrast, the Planck collaboration's findings in 2015 indicated $\omega_0 = -1.006 \pm 0.045$ \cite{Ade/2015}, and in 2018, it reported $\omega_0 = -1.028 \pm 0.032$ \cite{Planck_2020}. In addition, for the constrained values of the model parameters, we determine the present value of the EoS parameter as $\omega_0 = -1.07$, $\omega_0 = -1.19$, and $\omega_0 = -1.12$, respectively \cite{Gong/2007,Novosyadlyj/2012,Kumar/2014}. The observational data referenced serve as constraints on the EoS parameter, providing useful benchmarks. However, these constraints do not directly imply the presence of quintessence or phantom DE in our model, but instead highlight how modified gravity effects can lead to an evolving EoS similar to that of DE models.

\subsection{Energy conditions}

In GR, energy conditions are a set of constraints imposed on the stress-energy tensor $T_{\mu \nu}$ to ensure physically reasonable matter and energy distributions. These conditions help in understanding the behavior of matter and energy under various circumstances, especially in the context of gravitational collapse, black holes, and cosmological models. Here, the primary role of these energy conditions is to verify the accelerated expansion of the universe \cite{Barcelo,Moraes,Visser}. These conditions originate from the well-established Raychaudhuri equations, which describe the behavior of geodesic congruences (families) in a given spacetime, illustrating the focusing of geodesic flows due to gravity \cite{Raychaudhuri/1955}. The Raychaudhuri equation has different forms depending on the type of geodesics considered, whether timelike or null \cite{Nojiri2, Ehlers}:

1. Timelike geodesics:
\begin{equation}
\frac{d\theta}{d\tau} = -\frac{1}{3} \theta^2 - \sigma_{\mu\nu} \sigma^{\mu\nu} + \omega_{\mu\nu} \omega^{\mu\nu} - R_{\mu\nu} u^\mu u^\nu
\end{equation}
where $\theta$ is the expansion scalar, $\sigma_{\mu\nu}$ is the shear tensor, $\omega_{\mu\nu}$ is the vorticity tensor, $R_{\mu\nu}$ is the Ricci tensor, and $u^\mu$ is the timelike vector tangent to the geodesics.

2. Null geodesics:
\begin{equation}
\frac{d\theta}{d\lambda} = -\frac{1}{2} \theta^2 - \sigma_{\mu\nu} \sigma^{\mu\nu} + \omega_{\mu\nu} \omega^{\mu\nu} - R_{\mu\nu} k^\mu k^\nu    
\end{equation}
where $\theta$ is the expansion scalar for null geodesics and $k^\mu$ is the null vector tangent to the geodesics.

The Raychaudhuri equations are fundamental in understanding gravitational focusing and the formation of singularities in spacetime. They show how the expansion, shear, and rotation of geodesic congruences evolve and are influenced by the curvature of spacetime. The main energy conditions are:

1. Null energy condition (NEC):
\begin{equation}
T_{\mu\nu} k^\mu k^\nu \geq 0
\end{equation}
for any null vector $k^\mu$. This condition implies that the energy density as seen by a light-like observer is non-negative. This results in the form $\rho + p \geq 0$. When $ \omega < -1 $, it violates the NEC, which in turn leads to the violation of the second law of thermodynamics.

2. Weak energy condition (WEC):
\begin{equation}
   T_{\mu\nu} u^\mu u^\nu \geq 0
\end{equation}
for any timelike vector $ u^\mu $. This condition ensures that the energy density measured by any observer is non-negative. In addition, the NEC must be satisfied. This condition for energy density simplifies to $\rho \geq 0$ and $\rho + p \geq 0$.

3. Dominant energy condition (DEC):
\begin{equation}
   T_{\mu\nu} u^\mu u^\nu \geq 0 \quad \text{and} \quad T^{\mu\nu} u_\nu \, \text{is a non-spacelike vector}
\end{equation}
for any timelike vector \( u^\mu \). This condition ensures that energy density is non-negative and energy flux is non-spacelike, meaning energy cannot flow faster than light. This condition leads to $\rho \geq 0 $ and $ \rho \pm p \geq 0 $.

4. Strong energy condition (SEC):
\begin{equation}
   \left( T_{\mu\nu} - \frac{1}{2} T g_{\mu\nu} \right) u^\mu u^\nu \geq 0
\end{equation}
for any timelike vector $ u^\mu $, where $ T = T^\alpha_\alpha $ is the trace of the stress-energy tensor. This condition implies that gravity is always attractive, which leads to $ \rho + p \geq 0 $ and $ \rho + 3p \geq 0 $.

\begin{figure}[h]
\centerline{\includegraphics[scale=0.7]{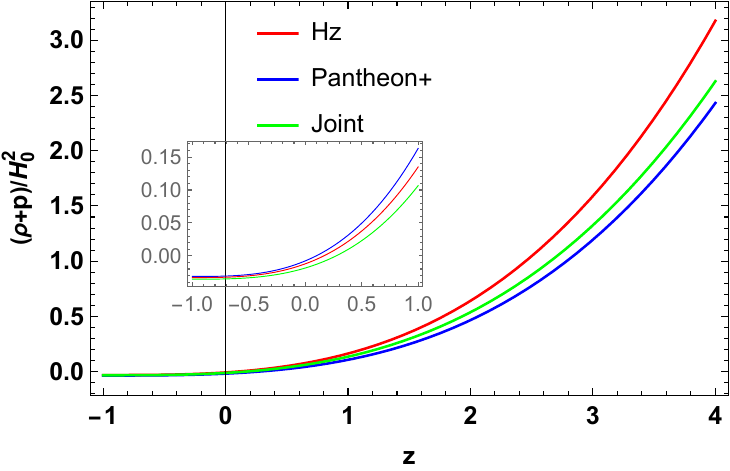}}
\caption{Variation of the NEC as a function of redshift $z$ with $b= -0.1$ and $n = 0.087$ \cite{Bekkhozhayev}.}
\label{F_NEC}
\end{figure}

\begin{figure}[h]
\centerline{\includegraphics[scale=0.7]{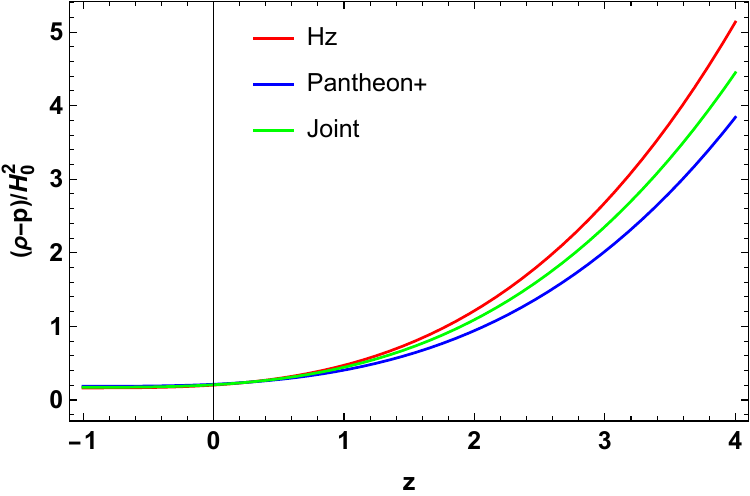}}
\caption{Variation of the DEC as a function of redshift $z$ with $b= -0.1$ and $n = 0.087$ \cite{Bekkhozhayev}.}
\label{F_DEC}
\end{figure}

\begin{figure}[h]
\centerline{\includegraphics[scale=0.7]{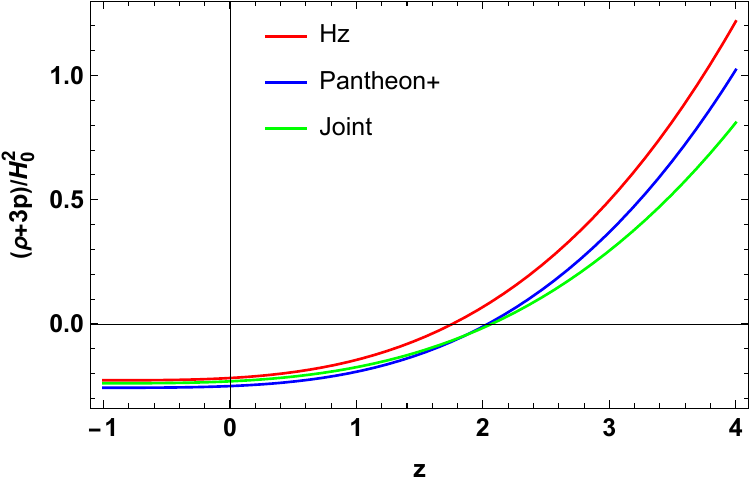}}
\caption{Variation of the SEC as a function of redshift $z$ with $b= -0.1$ and $n = 0.087$ \cite{Bekkhozhayev}.}
\label{F_SEC}
\end{figure}

The above energy conditions indicate that the violation of the NEC leads to the violation of the other energy conditions. This signifies a depletion of energy density as the universe expands. Furthermore, the violation of the SEC represents the acceleration of the universe. Figs. \ref{F_NEC}, \ref{F_DEC}, and \ref{F_SEC} show the evolution of the energy conditions with respect to redshift. From these figures, we observe that $ \rho + p \leq 0 $ and $ \rho + 3p \leq 0 $, indicating the violation of both the NEC and the SEC at present ($z=0$) and in the future ($z \rightarrow -1$). In addition, the figures show that $ \rho - p \geq 0 $, demonstrating that the DEC is satisfied. Therefore, the violation of the NEC and SEC, along with $ \omega < -1 $, suggests the phantom-like behavior of the universe, which is associated with its accelerated expansion.

\section{Concluding remarks}
\label{sec7}

The accelerating expansion of the universe remains one of the most compelling phenomena in modern cosmology. Taking into account anomalies observed in the CMB \cite{Bennett_1996}, we have explored the implications of $f(Q, T)$ gravity in a Bianchi type-I spacetime, which is characterized by spatial homogeneity and anisotropy, to study cosmic acceleration. Our approach, using a linear combination of the non-metricity $Q$ and the trace of the energy-momentum tensor $T$ in the form $f(Q,T) = Q + bT$, has allowed us to investigate the evolution of the universe in a modified gravity framework.

Through the parametrization of the deceleration parameter and the derivation of the Hubble solution, we imposed these conditions in the Friedmann equations for $f(Q,T)$ gravity. Employing a Bayesian approach, we estimated the best-fit values of the model parameters using MCMC sampling, constrained by $H(z)$, Pantheon+ samples, and combined observational data. The corresponding $1-\sigma$ and $2-\sigma$ contour plots for the $H(z)$ data, Pantheon+ sample, and the combined observational data are displayed in Figs. \ref{F_CC}, \ref{F_SNe}, and \ref{F_CC+SNe}, respectively. The obtained best fit values are: $H_0=67.8^{+1.3}_{-1.3}$, $\alpha=1.53^{+0.43}_{-0.40}$, and $\beta=2.2^{+1.5}_{-1.5}$ for $H(z)$ data; $H_0=72.0^{+1.4}_{-1.4}$, $\alpha=1.56^{+0.14}_{-0.13}$, and $\beta=3.55^{+1.0}_{-0.90}$ for Pantheon+ data; $H_0=68.1^{+1.2}_{-1.2}$, $\alpha=1.508^{+0.059}_{-0.054}$, and $\beta=2.57^{+0.58}_{-0.53}$ for the combined data. Our results indicate that the model successfully describes the transition from a decelerated epoch to an accelerated de-Sitter-like phase, with the transition redshift $z_t$ varying depending on the dataset. Specifically, the transition redshift is found to be $z_t = 0.61$ for the $H(z)$ data, $z_t = 0.85$ for the Pantheon+ samples, and $z_t = 0.72$ for the combined data. The present value of the deceleration parameter is determined to be $q_0 = -0.52$, $q_0 =-0.65$, and $q_0 =-0.57$, respectively. 

In addition, our analysis shows that the energy density starts with a high initial value and decreases over time, approaching zero for the present and future, indicating the continuous expansion of the universe. Moreover, the pressure initially exhibits large negative values, gradually transitioning to smaller negative values, driven by the modified gravity effects rather than a DE component. One of the key results of our analysis is the dynamic nature of the equation of state (EoS) parameter $\omega$, which evolves from quintessence-like behavior ($\omega > -1$) to the phantom regime ($\omega < -1$), a hallmark of models that cross the phantom divide. This behavior is driven entirely by the geometric contributions of the modified gravity model, without invoking a separate DE component. Furthermore, the present value of the EoS parameter is found to be $\omega$ is $\omega_0 = -1.07$, $\omega_0 = -1.19$, and $\omega_0 = -1.12$ for the different datasets, respectively. Lastly, we analyzed the evolution of the energy conditions. We observed the violation of both the NEC and the SEC at present and in the future. However, the DEC is satisfied. Therefore, the violation of these energy conditions implies phantom-like behavior and the accelerated expansion of the universe. Lymperis \cite{Lymperis} explored phantom DE within $f(Q)$ gravity, showing how modified gravity can drive cosmic acceleration. Our findings, particularly the transition of the EoS into the phantom regime, align with this work, further supporting the role of $f(Q, T)$ gravity in explaining late-time acceleration without invoking an explicit DE component.

Future research directions in $f(Q, T)$ gravity could involve exploring more complex functional forms of $ f(Q, T)$, such as $f(Q, T) = Q^n + b T$ \cite{Loo_2023}, to account for a broader range of cosmological phenomena. In addition, addressing standing issues such as the detailed stability of solutions, the investigation of perturbation theory \cite{Najera}, and the connection to quantum gravity could deepen our understanding of the model's predictions. In anisotropic cosmologies, extending this framework to include interactions between different cosmic components or exploring the impact of inhomogeneities could provide new insights into the early and late universe's structure.

\section*{Data Availability Statement}
There are no new data associated with this article.

\end{document}